\documentclass[twoside]{article}

\def\be{\begin{equation}}
\def\ee{\end{equation}}

\def\BibTeX{{\rm B\kern-.05em{\sc i\kern-.025em b}\kern-.08em
            T\kern-.1667em\lower.7ex\hbox{E}\kern-.125emX}}

\usepackage{graphics,epsfig}
\begin{document}
\sloppy
\twocolumn[{
\vspace*{1.7cm}   
\begin{center}
{\large\bf STUDY OF EVENT-BY-EVENT FLUCTUATIONS IN}
{\large\bf HEAVY ION COLLISIONS}\\

{\small B.~Tom\'a\v{s}ik$^{a,b}$ (boris.tomasik@umb.sk),
I.~Melo$^c$ (melo@fyzika.uniza.sk), 
M.~Gintner$^{a,c}$ (gintner@fyzika.uniza.sk), S.~Kor\'ony$^a$ (samuel.korony@umb.sk), \\
$^a$ Univerzita Mateja Bela, Bansk\'a Bystrica, Slovakia, $^b$ \v{C}esk\'e
vysok\'e u\v{c}en\'{\i} technick\'e,
Praha, Czech Republic, 
$^c$ \v{Z}ilinsk\'a Univerzita,
\v{Z}ilina, Slovakia,}\\

\end{center}
\vspace*{1ex}

{\bf ABSTRACT.} We propose Kolmogorov-Smirnov test as a means for recognising event-by-event fluctuations of
rapidity distributions in relativistic heavy ion collisions. Such fluctuations may be induced
by the spinodal decomposition of the rapidly expanding system during the 1st order quark-hadron
phase transition.\\ \\
}]

\section{INTRODUCTION}

In heavy ion collisions matter in state with deconfined quarks (quark-gluon plasma) 
may be produced 
depending on the collision energy. The deconfined matter expands and cools down
very quickly, possibly reaching the line of the 1st order phase transition between
the quark phase and the hadronic phase. During this transition the system supercools
and if the expansion rate is fast enough it may go through the so-called spinodal 
decomposition - the fragmentation of quark-gluon plasma into pieces of 
characteristic size \cite{mfrag,Tomasik,Scavenius}. 
These fragments will eventually hadronize, emitting particles with 
a unique rapidity spectrum which reflects a unique nature of fragmentation of a given event.
Rapidity spectra thus fluctuate non-statistically on event-by-event basis.

We propose a method for the study of these fluctuations based on 
Kolmogorov-Smirnov (KS) test. KS test addresses the question
if two empirical distributions $F_{n_1}(X), F_{n_2}(X)$ are generated 
from {\it the same}
underlying probability distribution.
The maximum difference $D$ between $F_{n_1}(X), F_{n_2}(X)$ is distributed
according to 
\begin{equation}
H = \lim_{n_1,n_2 \rightarrow \infty} P(\sqrt{n} D < t) = 
\sum_{k=-\infty}^{k=\infty}(-1)^k \exp{(-2 k^2 t^2)}
\end{equation}
with $n=n_1 n_2/(n_1+n_2)$.

\section{RESULTS}

Below we calculate quantity $Q = 1 - H$ for rapidity spectra 
(empirical distributions $F_{n_1}(X), F_{n_2}(X)$) of random pairs of simulated 
events where rapidities were first generated according to test (Gaussian) distributions 
and then for
pairs of events generated with our Monte Carlo droplet generator which simulates particle 
emission from the
fragments emerging from spinodal decomposition.  
If rapidity spectrum for each event is generated from the same underlying distribution, we expect
an equal number of event pairs at any $Q$ - a uniform, constant line histogram (we expect this in 
the case when no fragments are present). 
For events drawn from different 
underlying distributions the pairs group at low $Q$ values, indicating non-statistical fluctuations. 

This point is illustrated in Fig. 1 where we show results of KS test on 
data randomly generated from two different Gaussian distributions:
{\bf a)} half of $10^5$ events was generated according to the distribution $(\mu, \sigma ) = (1,0.1)$, half
according to $(\mu, \sigma ) = (2,0.1)$,  
{\bf b)} half according to $(\mu, \sigma ) = (1.99,0.1)$, half
according to $(\mu, \sigma ) = (2,0.1)$. Multiplicities ranged in both cases from $8$ to $535$.
The Q value was calculated for $10^5$ random pairs of these events and the histogram of the 
number of pairs as a function of the corresponding Q value was plotted.

\setlength{\unitlength}{1.0cm}
\begin{picture}(9,11.5)
\put(-0.4,6){\begin{picture}(9,5.5)
\epsfig{file=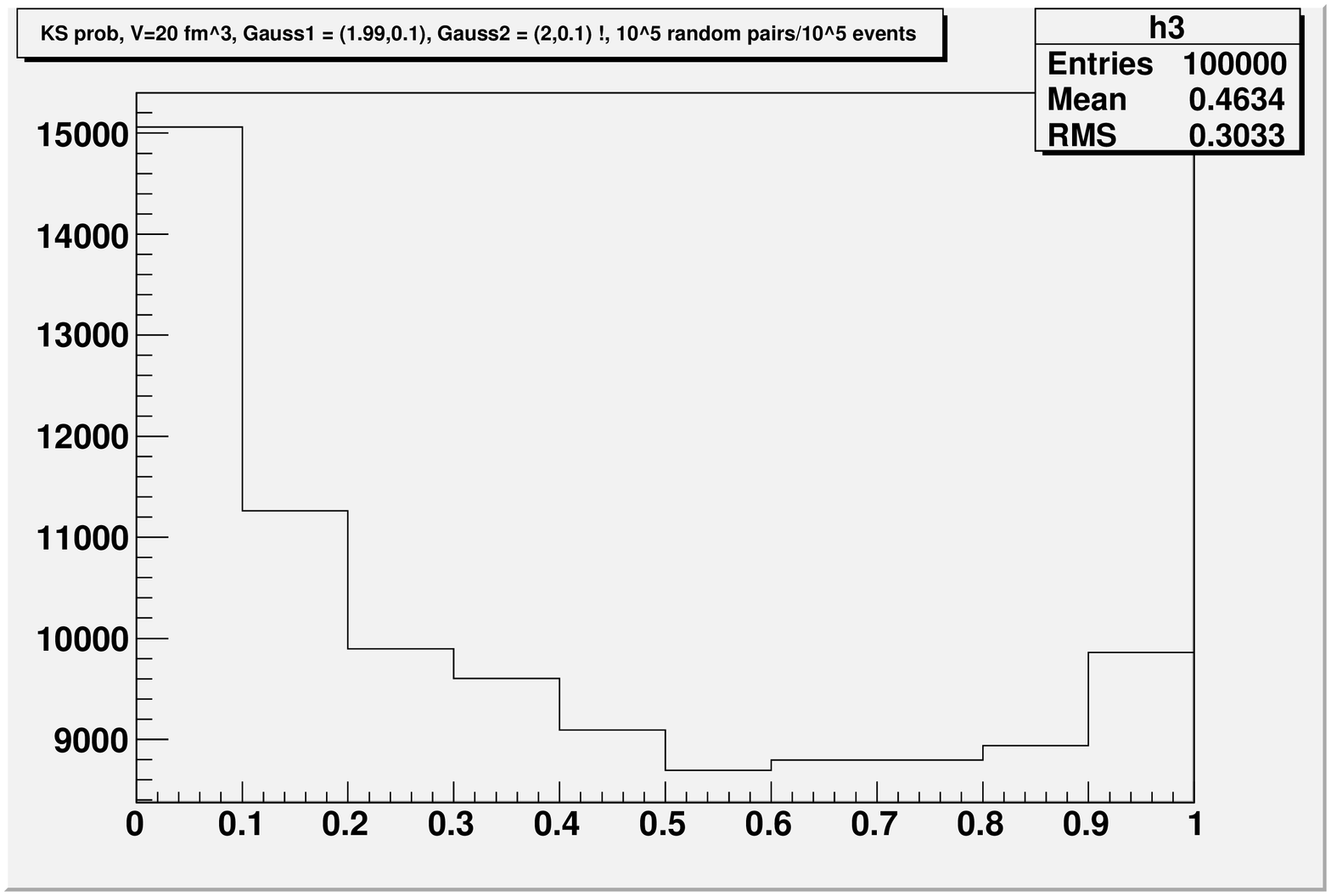,height=90mm,width=55mm,clip=,silent=,angle=-90,
bbllx=2.5cm,bblly=1cm,bburx=20cm,bbury=30cm}
\put(-3.4,-1.6){(b)}
\put(-2.1,-5.45){{\small Q}}
\end{picture}}
\put(-0.4,11.4){\begin{picture}(9,5.5)
\epsfig{file=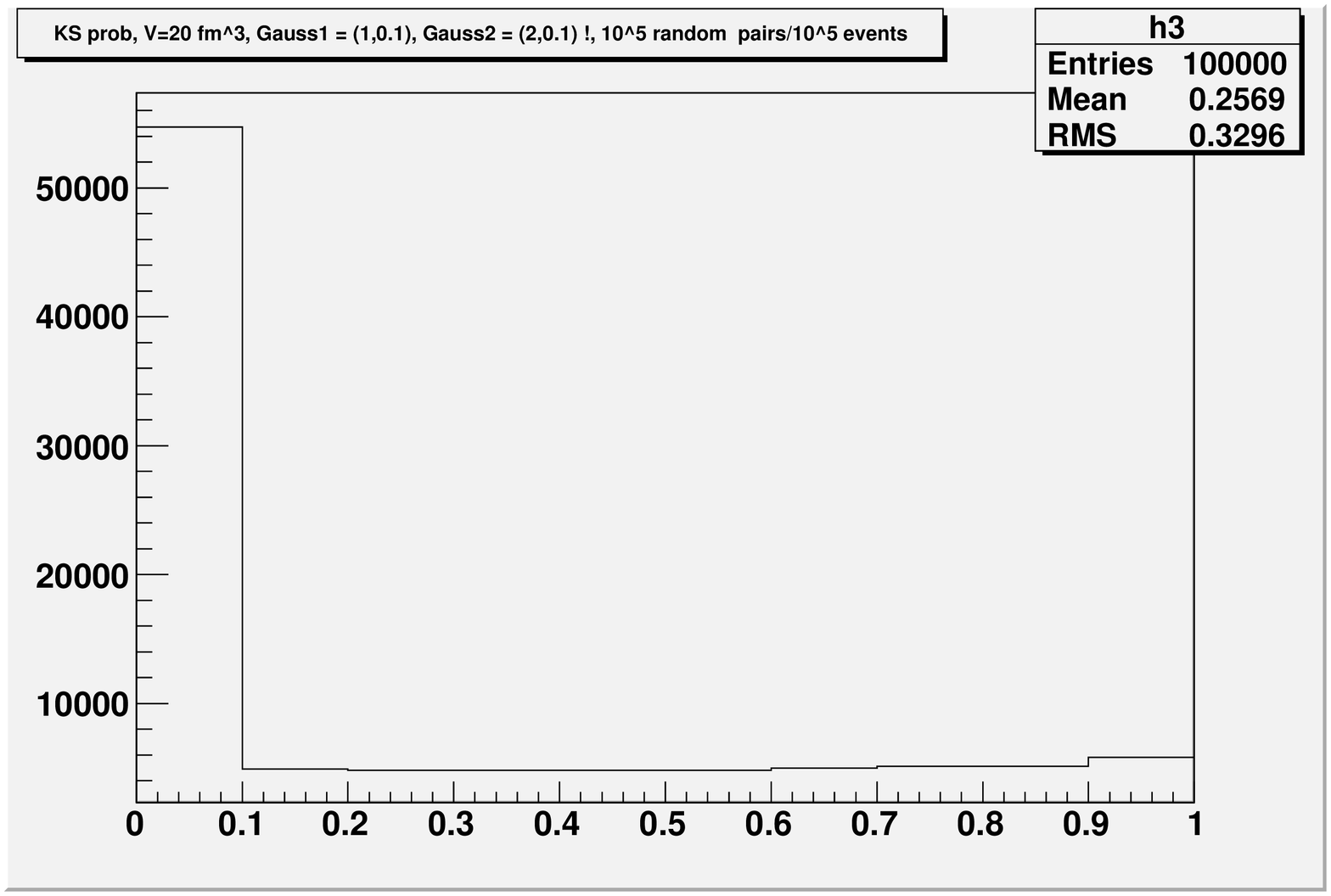,height=90mm,width=55mm,clip=,silent=,angle=-90,
bbllx=2.5cm,bblly=1cm,bburx=20cm,bbury=30cm}
\put(-3.4,-1.6){(a)}
\put(-2.1,-5.45){{\small Q}}
\end{picture}}
\end{picture}

\vspace{-2mm} 

\noindent Fig. 1. KS test for data  from two different Gaussian distributions:
{\bf a)}  $(\mu, \sigma ) = (1,0.1)$ and $(\mu, \sigma ) = (2,0.1)$,  
{\bf b)}  $(\mu, \sigma ) = (1.99,0.1)$ and $(\mu, \sigma ) = (2,0.1)$.
\bigskip

We can see in Fig. 1a that half of pairs is distributed uniformly (5000 pairs in each bin) 
and half (50 000) is concentrated in the first bin.
The first half consists of events generated from the same distribution (like events) 
which form a constant line, the second half
consists of events from opposite distributions (unlike events) which group at small Q values. 
Note that the statistical fluctuation is $\pm \sqrt{N}$. The deviation
from the constant line at $N=10000$ above $\sqrt{N} = 100$ represents non-statistical
fluctuation.

Fig. 1b shows the power of the KS test. Although the two Gauss distributions differ 
by just 0.01 in their mean value ($\sigma = 0.1$), the
KS test clearly recognized that the events were not drawn from a single distribution.


After the test of KS test on two Gaussian distributions, we applied the KS test
on data generated with the Monte Carlo droplet generator. This generator simulates particle
rapidity spectra emerging from the fragments (droplets) originating in the spinodal decomposition.
Droplets are generated from a blast-wave source with tunable parameters. 
In Fig. 2 we show results of KS test on droplets of size a) $50$ fm$^3$ and b) $10$ fm$^3$ 
at temperature $T= 175$ MeV. As before, we used a sample of $10^5$ event pairs randomly chosen 
out of $10^5$ events. All particles come from 
the droplets.
Non-statistical fluctuations are clearly visible in both cases 
and grow with the droplet size.
 

\setlength{\unitlength}{1.0cm}
\begin{picture}(9,11.5)
\put(-0.4,6){\begin{picture}(9,5.5)
\epsfig{file=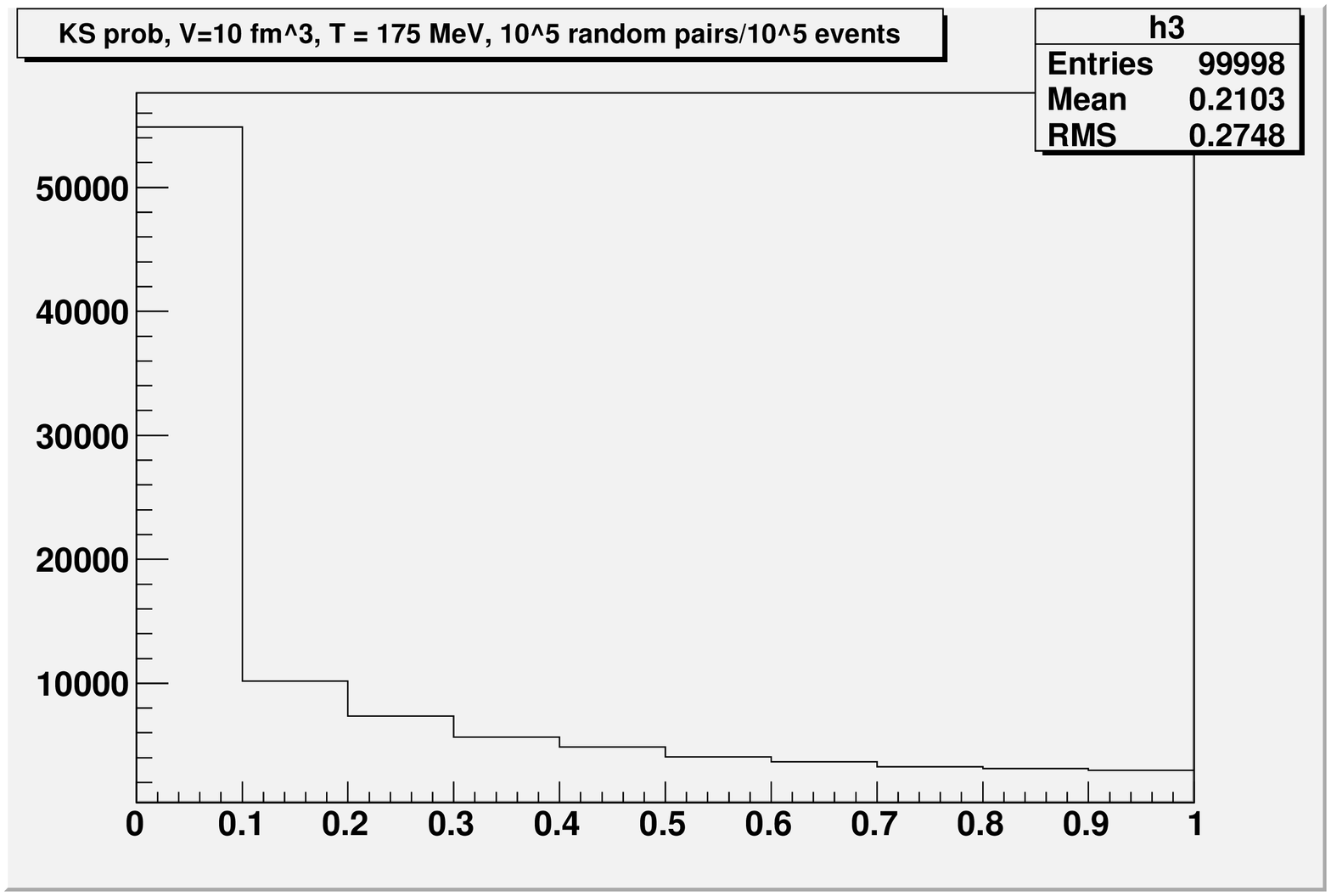,height=90mm,width=55mm,clip=,silent=,angle=-90,
bbllx=2.5cm,bblly=1cm,bburx=20cm,bbury=30cm}
\put(-3.4,-1.6){(b)}
\end{picture}}
\put(-0.4,11.4){\begin{picture}(9,5.5)
\epsfig{file=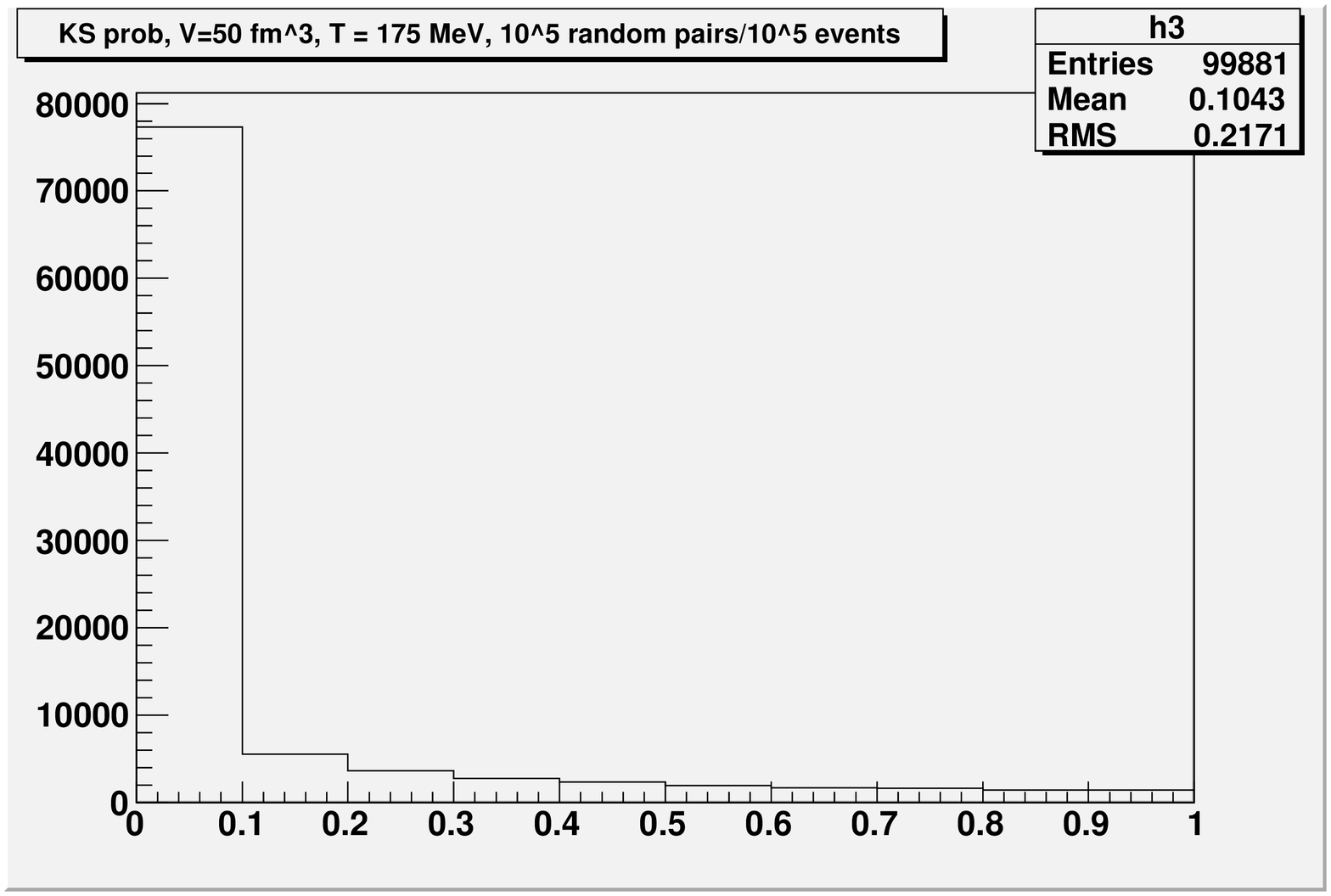,height=90mm,width=55mm,clip=,silent=,angle=-90,
bbllx=2.5cm,bblly=1cm,bburx=20cm,bbury=30cm}
\put(-3.4,-1.6){(a)}
\end{picture}}
\end{picture}

\vspace{-2mm} 

\noindent Fig. 2. KS test on droplets of size a) $50$ fm$^3$ and b) $10$~fm$^3$ 
at temperature $T= 175$ MeV.

\bigskip
Finally, we fix the droplet size at $10$~fm$^3$ ($T = 175$ MeV) and vary the percentage of particles
originating in the droplets.  In Fig. 3a $20\%$ of particles comes from the droplets 
as compared with $60\%$ in Fig. 3b. Again the sample was $10^5$ event pairs chosen randomly from
$10^5$ simulated events. The non-statistical fluctuations are present even for the $20\%$ case.
They vanish as expected for the $0\%$ case (not shown).

\setlength{\unitlength}{1.0cm}
\begin{picture}(9,11.5)
\put(-0.4,6){\begin{picture}(9,5.5)
\epsfig{file=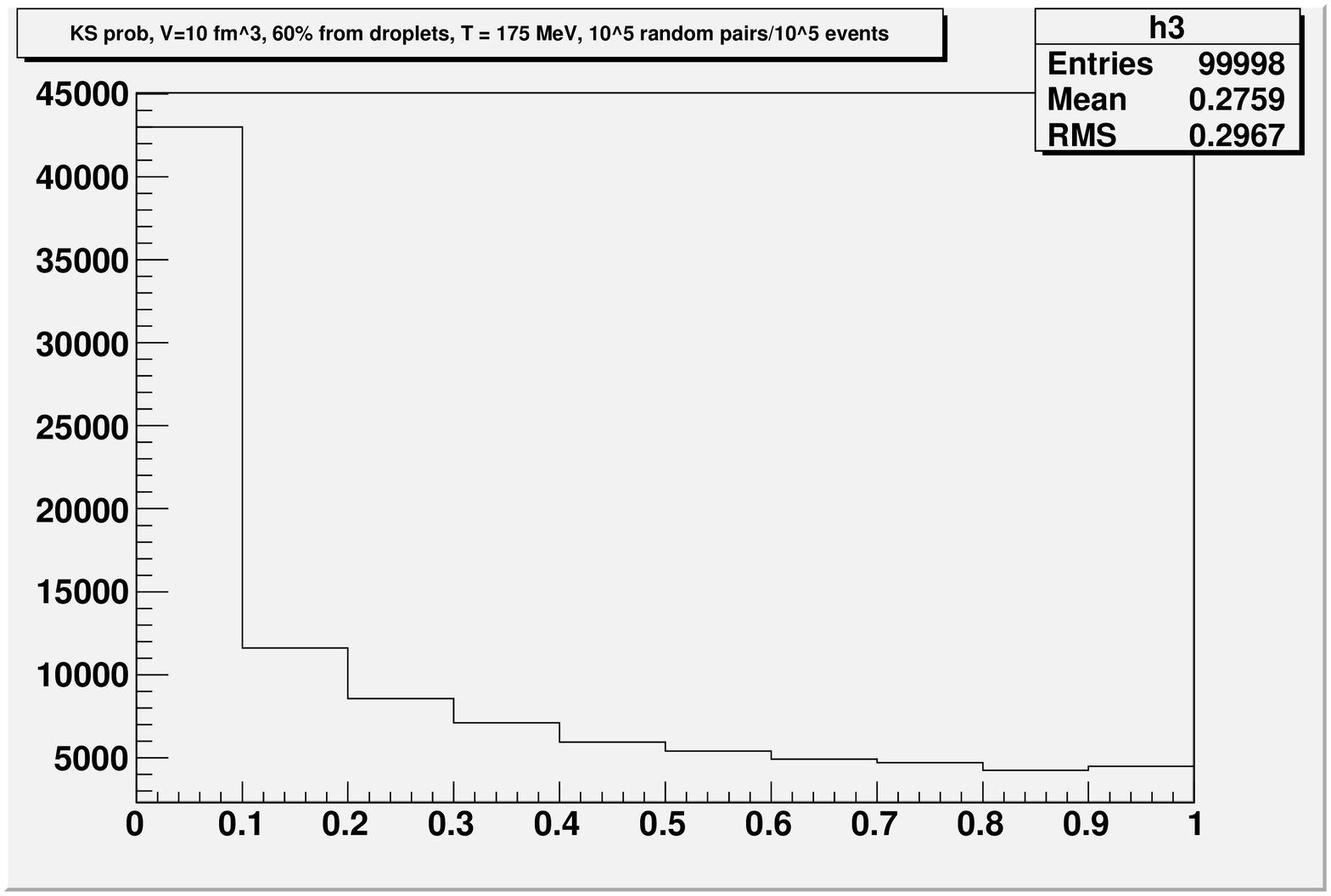,height=90mm,width=55mm,clip=,silent=,angle=-90,
bbllx=2.5cm,bblly=1cm,bburx=20cm,bbury=30cm}
\put(-3.4,-1.6){(b)}
\end{picture}}
\put(-0.4,11.4){\begin{picture}(9,5.5)
\epsfig{file=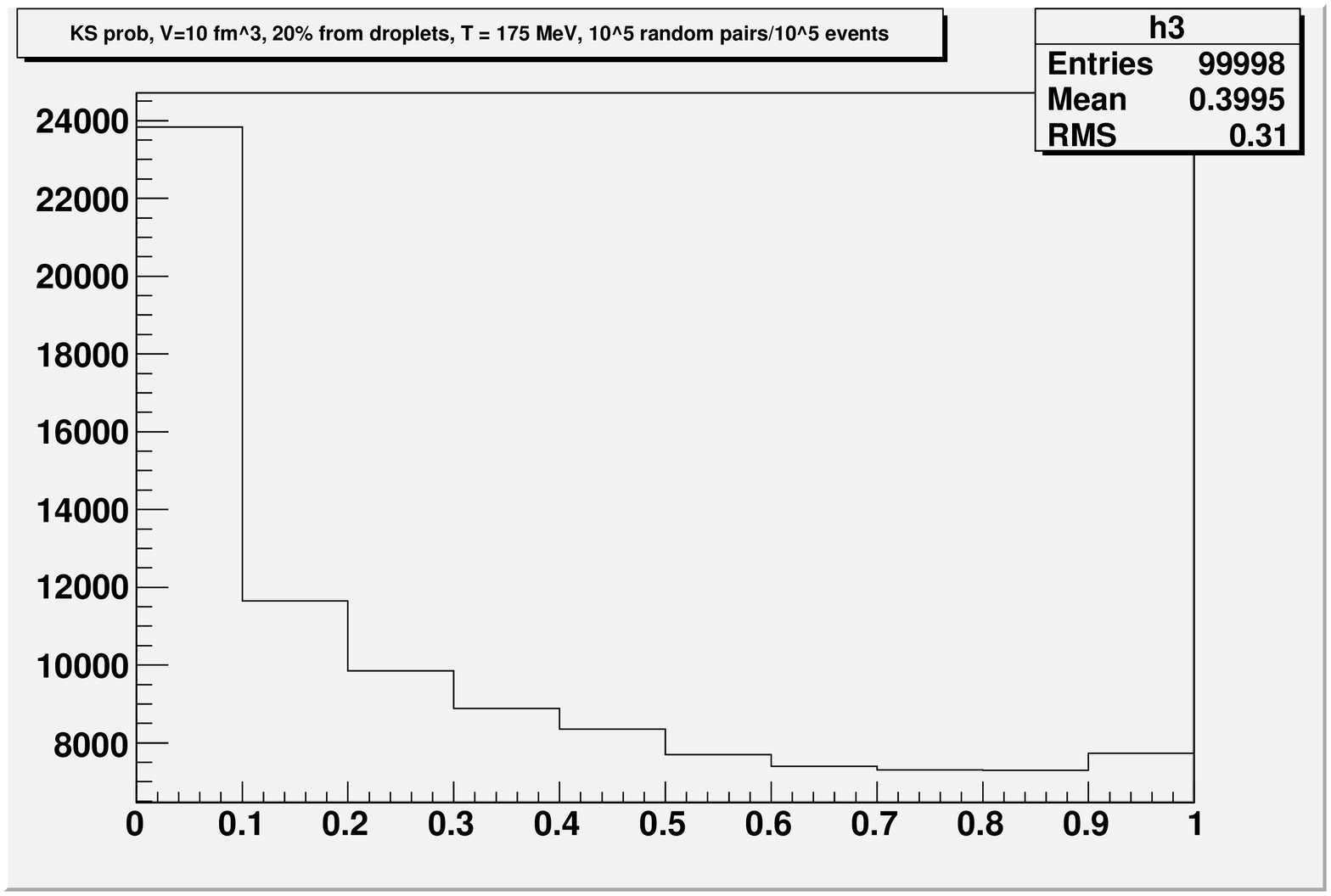,height=90mm,width=55mm,clip=,silent=,angle=-90,
bbllx=2.5cm,bblly=1cm,bburx=20cm,bbury=30cm}
\put(-3.4,-1.6){(a)}
\end{picture}}
\end{picture}

\vspace{-2mm} 

\noindent Fig. 3. KS test for the case when only a fraction of particles comes from the droplets:
a) $20\%$  and b) $60\%$.  

\bigskip

\section{CONCLUSIONS}

\noindent 
We studied event-by-event fluctuations of rapidity distributions by means of the KS test.
We have shown that the KS test can recognize whether or not the rapidity spectra are drawn from 
the same underlying distribution. If spinodal decomposition occurs during the 1st order 
quark-hadron phase transition, the resulting droplets lead to many underlying distributions and thus
to event-by-event fluctuations. As a next step, we plan to apply this method on data
by NA49 collaboration. Of course, if the fluctuations are indeed there, it does not follow automatically that
they are caused by the droplets.   

This research has been supported by VEGA 1/4012/07.


\begin{thebibliography}{99}

\leftskip=-5pt \vspace{-0.3truecm}
\bibitem{Tomasik} B. Tom\'a\v{s}ik {\it et al.}, contribution to this conference, 
arXiv:~0711.4932 [nucl-th].
\bibitem{Scavenius} O. Scavenius {\it et al.}, Phys. Rev. D {\bf 63}, 116003 (1993).
\bibitem{mfrag}
I.~N.~Mishustin,
Phys.\ Rev.\ Lett.\  {\bf 82}, 4779 (1999).

\end{thebibliography}
\end{document}